\documentclass[prl,twocolumn,aps,showpacs]{revtex4}
                                                                                                         
\usepackage{t1enc}
\usepackage[final]{graphicx}
\usepackage{graphicx}
\usepackage{epsfig}

\begin{document}
\preprint{APS/123-QED}

\title{Spin-glass behavior in the random-anisotropy Heisenberg model}         

\author{Orlando V. Billoni}
\email{billoni@famaf.unc.edu.ar}
\altaffiliation{Postdoc fellow of CONICET, Argentina}
\affiliation{ Facultad de Matem\'atica, Astronom\'\i a y F\'\i sica,
         Universidad Nacional de C\'ordoba,
         Ciudad Universitaria, 5000 C\'ordoba, Argentina}

\author{Sergio A. Cannas}
\altaffiliation{Member of CONICET, Argentina}
\affiliation{ Facultad de Matem\'atica, Astronom\'\i a y F\'\i sica,
         Universidad Nacional de C\'ordoba,
         Ciudad Universitaria, 5000 C\'ordoba, Argentina}

\author{Francisco A. Tamarit}
\altaffiliation{Member of CONICET, Argentina}
\affiliation{ Facultad de Matem\'atica, Astronom\'\i a y F\'\i sica,
         Universidad Nacional de C\'ordoba,
         Ciudad Universitaria, 5000 C\'ordoba, Argentina}

\date{\today}                                           
                                                                                                         
\begin{abstract}
We perform Monte Carlo simulations in a random anisotropy magnet at a intermediate 
exchange to anisotropy ratio. We focus on the out of equilibrium relaxation after a sudden quenching
in the low temperature phase, well below the freezing one. By analyzing both the
aging dynamics and the violation of the Fluctuation Dissipation relation we found
strong evidence of a spin--glass like behavior. In fact, our results are qualitatively
similar to those experimentally obtained recently in a Heisenberg-like real spin glass.
\end{abstract}

\maketitle
     
\section{Introduction}
Random magnetic anisotropy seems to be a fundamental ingredient for any realistic
description of amorphous materials, which are systems of both practical and theorical
relevancy.
For the particular case of amorphous alloys \cite{Hellman98PRB,Pickart74PRL} of
non-S-state rare earths and transition metals (RE-TM), such as
Tb$_x$Fe$_{1-x}$, the three dimensional classical Heisenberg model with random
uniaxial single-site anisotropy (RAM) is considered to be the proper model for studying
both their thermodynamic and dynamic properties.
This model was introduced by Harris, Plischke, and Zuckermann, \cite{Harris73PRL}
who performed a mean-field calculation and found a ferromagnetic (FM) phase at
low temperature. Later on,  Pelcovits, Pytte and Rudnick  \cite{Pelcovits78PRL} claimed, using
an argument similar to that used by Imry and Ma \cite{Imry75PRL} for the random-field
case, that such a FM phase is not stable in three dimensional RAM model, for any
finite value of the anisotropy. Since then, the nature of the ordered phase at low
temperatures and its dependence on the degree of anisotropy is a subject of controversy.
Recently, Itakura \cite{Itakura03PRB} proposed, by using Monte Carlo simulations and
referring to former works in the literature, a schematic phase diagram for the RAM where
different kinds of order can be found, depending both on the temperature and the
anisotropy strength of the system.

When an amorphous material is cooled, it can eventually get blocked at certain 
temperature $T_f$, at which the magnetic moments freeze pointing in random directions.
The value of $T_f$ strongly depends on the degree of anisotropy,
the strength of the interactions between domains and the cooling rate. 
It is worth here to stress that this {\em freezing process} is a
dynamical phenomenon which can not be associated to any true thermodynamical phase 
transition. In particular, this phenomenology has been reported, during the last 
years, in the study of hard magnetic amorphous alloys \cite{Inoue96JIM,Billoni03JMMM}, 
 which has been also successfully simulated using a slightly modified version of the RAM \cite{Wang01PRB}. 
On the other hand, in experimental spin  glasses \cite{Chamberlin82PRB},  zero field 
cooling  (ZFC) and field cooling (FC) curves of magnetization versus temperature 
are useful to estimate the characteristic freezing temperatures of the systems. 

Most of the numerical effort  in the study of the RAM model concerns the case of strong or
infinite anisotropy. In this case the model seems to present a low temperature spin glass like
phase (usually called speromagnetic \cite{Coey78JAP}). On the other hand, in the weak anisotropy limit, the 
system tends to order locally in a ferromagnetic state \cite{Chudnovsky86PRB} (also called 
asperomagnetic).  When increased to intermediate values,the anisotropy seems to destroy
the asperomagnetic ferromagnetic state, and the systems orders in the so called {\em 
correlated speromagnetic} or {\em correlated spin--glass}. It has been recently 
verified in Ref.\cite{Itakura03PRB}, by means of a very extensive numerical simulation,
that the competition between exchange and anisotropy gives place to a 
quasi--long--range order (QLRO) low temperature phase characterized by frozen 
power--law spin--spin correlations. 
It is important here to remark that 
all the theoretical results ultimately confirm the observed lack of long range
ferromagnetic order observed experimentally in magnetic materials with isotropic
easy axis distribution \cite{Dudka04CM}. 

Although the RAM  model has deserve much attention during the last years, 
most of the works were concerned on its equilibrium properties as well as on
its magnetic behavior, paying little attention to its relaxation dynamics.
The main question we want to address in this work concerns the possible
existence of a spin--glass like dynamical  behavior associated with QLRO 
low temperature phase in the intermediate anisotropy regime.
We analyze, through Monte Carlo simulations, the out of equilibrium relaxation 
of the three dimensional RAM model defined on a cubic lattice. In particular,
we report results obtained in the low temperature phase (well below $T_f$) and
for intermediate values of the anisotropy to exchange ratio. 

\section{The model}
The system is ruled by the following classical Heisenberg Hamiltonian:
\begin{equation}
H=-J \sum_{<i,j>} \vec{S}_{i} \cdot \vec{S}_{j} - D \sum_{i} (\vec{n}_i 
\cdot \vec{S}_i)^2 - 
\vec{H}_i \cdot \vec{S}_i
\end{equation}
where $D$ and $J>0$ are the anisotropy and the exchange strength, respectively, and
$\vec{H}_i$ is an external field acting on the site $i$. The spin variable $\vec{S}_i$ 
is a three components unit vector associated to the $i$--th node of the lattice and the 
first sum runs over all nearest-neighbor pairs of spins. $\vec{n}_i$ is unit 
random vector that defines the local easy axis direction of the anisotropy
at site $i$.  These easy axis are quenched variables chosen from a
isotropic distribution on the unit sphere.
The simulations were performed in a system of $N=L^3$ spins ($L=15$),
using a Monte Carlo Metropolis algorithm \cite{GarciaOtero99JAP}  with periodic
boundary conditions  (in fact, we have compare different system sizes up to $L=20$, confirming
that for $L=15$ finite size effects become very small). 
The ratio between the anisotropy and the exchange
strength was fixed at the value $D/J=3.5$, which is comparable to those
values observed in real amorphous material and clusters ferromagnetic alloys
\cite{Wang01PRB}. Following the ideas used in Ref. \cite{Wang01PRB}, at each spin 
actualization the direction of the spin is adjusted in such a way to maintain 
an acceptation rate close to 0.46.

\section{Results}
\begin{figure}
\includegraphics*[width=8cm,angle=0]{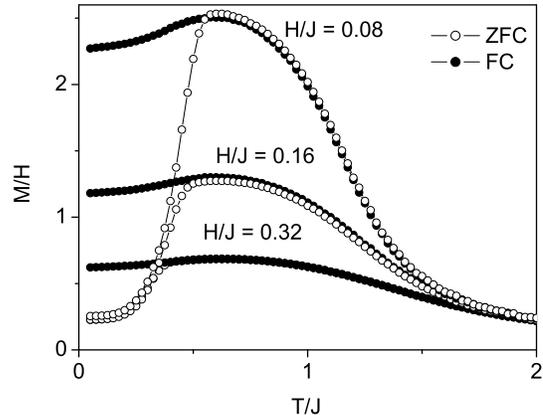}
\caption{\label{fig1} Magnetization divided by the
field $M/H$ (dc susceptibility) as a function of the rescaled temperature 
$T/J$ for different applied fields, both for the ZFC and FC processes.}
\end{figure}

The first step was to locate the {\em freezing temperature} $T_f$ for these particular
values of the parameters by looking for the temperature at which zero field 
cooling (ZFC) and field cooling (FC) curves split each other.
Fig.\ref{fig1} shows the magnetization divided by the field $M/H$ (dc susceptibility) 
as a function of $T/J$, both for the ZFC and FC processes.  The simulation protocol
used is the following: we performed 1000 Monte Carlo steps (MCS) at a given temperature
and at a  constant acceptance rate in order to equilibrate the system, after which, we used
other 1000 MCS  to get a time average of the magnetization before decreasing
the temperature. In all the simulations  presented in this
work we have used between 20 and 40 different realizations
of the disorder to average the results. 
We find that the freezing temperature $T_f/J$ is close to $0.5$. At very low temperature the 
dc susceptibility in the ZFC curves is constant and the magnetization almost 
zero, indicating  a speromagnetic {\em spin--glass like} order. In other
words, the spins are frozen into random orientations with average correlation over 
at most one lattice parameter (we have verified this behavior by calculating the
dependence of the spin-spin correlation function as function of the distance).
As temperature increases the dc susceptibility 
also increases, indicating that the system goes to an anisotropic asperomagnetic phase.
Here again, the same conclusion has been verified by analyzing the correlation 
function which stabilizes in a non zero value for distances larger than 
approximately four lattice units.
  
On the other hand, it can be seen that as $T\to 0$, $M_{FC} \to 0.5$, as 
expected in an asperomagnetic phase (and in agreement with other theoretical 
\cite{Chudnovsky86PRB} and numerical predictions \cite{Harris78JPFMP}).
Due to {\em local} character of this asperomagnetic phase, a sufficiently 
large system could eventually develop a local order without any global magnetization (QLRO), 
as in fact is observed in amorphous materials. 
Notice that both curves present, 
above $T_f$, an inflection point  which is very close to the Heisenberg 
critical temperature, indicating the loss of magnetic order and the entrance
in the paramagnetic phase (or superparamagnetic phase, in particulate systems).
It is worth here to remark that the FC curves display
a clear cusp, that closely resembles the experimental results
obtained in spin glasses, as for intance in AgMn \cite{Chamberlin82PRB},
confirming the random freezing of the spin orientations.

Disordered systems, when suddenly quenched down into the low temperature phase, suffer a
drastic slowing down of their relaxation dynamics. At the same time, a very
strong dependence on the history of the sample emerges, a phenomenon usually
called {\em aging}. In a real experiment, the simplest way for measuring aging
is by suddenly quenching the system without field into the ordered phase.
The system {\em ages} in this phase during certain waiting time $t_w$, at
which the field is switched off. The measurement of the relaxation of the magnetization  
then strongly depends on both, the age or waiting time $t_w$ and the time $t$ elapsed
since the field was turned off, indicating the loss of time translation invariance (TTI)
proper of any equilibrated state. In a computer simulation
the same effect can be visualized by measuring the two time auto correlation function
after a sudden quench from infinite temperature into the low temperature phase:
\begin{equation} C(t+t_w,t_w)= \left[ \frac{1}{N}\sum_{i}\left< \vec{S}_{i}(t_w) \cdot 
\vec{S}_{i}(t+t_w)\right> \right] 
\end{equation}
where $t_w$ is the time elapsed after the quenching and $[ \cdots ]$ means an average over the disorder.

\begin{figure}
\includegraphics*[width=8cm,angle=0]{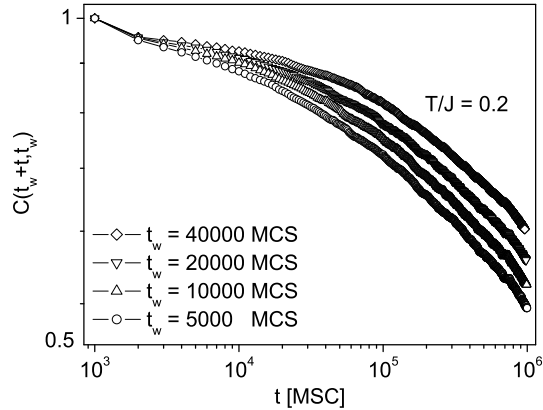}
\caption{\label{fig2} Correlation curves $C(t_w + t, t_w)$ as function of time $t$ 
(MCS) obtained at $T/J=0.2$, for $L=15$ and for different values of $t_w$.}
\end{figure}

In Fig. \ref{fig2} we present the curves of $C(t_w + t, t_w)$ obtained at
$T/J=0.2$, $L=15$ and for different values of $t_w$. The plot clearly confirms
the appearance of aging, characterized both by the loss of TTI and the
fact that the system decays slower as its age $t_w$ increases. 

Actually, aging is so ubiquitous in nature, that one can wander
whether it is useful or not to look for this phenomenology. But fortunately,
the peculiar dependence of $C(t_w+t,t_w)$ on $t$ and $t_w$ on a great
variety of systems (both theoretical and real systems) suggests the
existence of only a few {\em universality classes} associated to 
the out of equilibrium relaxation of the model, as occurs, for
instance, in critical phenomena and coarsening dynamics. 
All this indicates that, despite any microscopic difference between
different systems, the relaxation must be dominated by a few relevant
ingredients. 

\begin{figure}
\includegraphics*[width=8cm,angle=0]{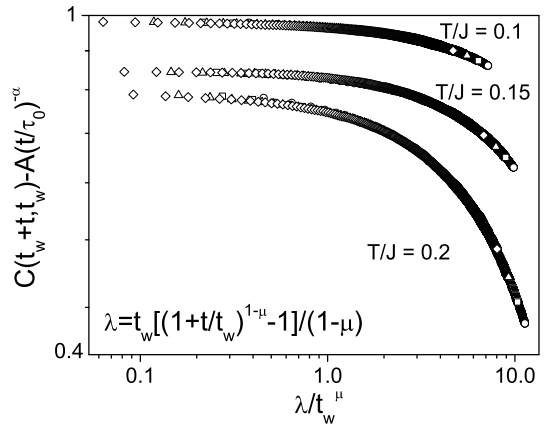}
\caption{\label{fig3} Rescaled correlations  curves for three different
values of $T/J$ ($0.1$, $0.15$ and $0.2$). Each curve we shown in the graph 
correspond to four correlation curves $C(t_w + t, t_w)$  with  $t_w=5000$, $10000$,
$20000$ and $40000$ (measured in MCS).}
\end{figure}

In order to classify the universality class, we looked for the best data
collapse of the curves obtained for different waiting times, checking different 
scaling forms. The best results we obtained are presented in Figure \ref{fig3}, 
where we followed the standard scaling procedure used in spin 
glasses materials (like for instance CdIn$_{0.3}$Cr$_{1.7}$S$_4$ 
\cite{Alba87JAP}). The two time autocorrelation function is assumed 
to have two different components
$C(t_w+t,t_w) = C_{st}(t) + C_{ag}(t_w+t,t_w)$,
the first one corresponding to an stationary part (independent of $t_w$) and the 
second part (the aging one) depending both on $t$ and $t_w$.
The stationary part is a power law function 
$A(\tau_0/t)^{\alpha}$, which is predominant at very 
small times. The aging part is a function of $\lambda(t,t_w)/t_w$,
where
\begin{equation}
\lambda = t_w [(1+t/t_w)^{1-\mu}-1]/[1-\mu]
\end{equation}

The curves presented in Fig.\ref{fig3} correspond to three different
values of $T/J$ ($0.1$, $0.15$ and $0.2$), all of them well below the freezing temperature of
the system.  In each curve we have collapsed the four different 
curves obtained for four different waiting times $t_w=5000$, $10000$,
$20000$ and $40000$ (measured in MCS).  Note the 
excellent superpositions obtained, which extends over
the complete time span simulated. It is worth to stress that this scaling law has 
been vastly used in the study of aging dynamics in real spin glass materials,
yielding excellent data collapses of the experimental results 
\cite{Vincent87JPC,Ladieu04CM,Alba87JAP,Vincent97SV}.

The set of fitting parameters obtained 
(shown in table \ref{tabla1}) are in a good agreement with those
obtained experimentally in real spin glasses \cite{Alba87JAP}. 
This behavior indicates that $C(t_w+t,t_w)$ scales as $t/t_w^{\mu}$ 
at short $t$  (since $\mu < 1$), while in the large $t$ limit and for
the values of $\mu$ obtained (which are all close to 1) its behaviors
is almost logarithmic on $t/t_w$ (as expected in a 
activated scenario \cite{Fisher88PRB}). 

\begin{table}
\caption{\label{tabla1}Fitting 
parameters of the scaling.}
\begin{ruledtabular}

\begin{tabular}{lccc}

\\[-3mm]
\vspace{1mm}
$T/J$    &  $A$ & $\alpha$ &   $\mu$ \\   
\hline \\[-3mm]
0.10     &  0.002   & 0.4    &  0.93       \\
0.15     &  0.012   & 0.06   &  0.86         \\
0.20     & 0.16     &  0.07  &  0.83           

\vspace{1mm}    \\
\end{tabular}
\end{ruledtabular}
\end{table}

Finally, we analyzed the Fluctuation Dissipation Relation (FDR), which can be expressed as  \cite{Cugliandolo93PRL}:

\begin{equation}
R(t_w+t,t_w) = \frac{X(t_w+t,t_w)}{3T} \frac{\partial C(t_w+t,t_w)}
{\partial t_w}
\end{equation}

where $R(t_w+t,t_w)= 1/N \; \sum_i \partial \left< S_i^z(t_w+t) \right>/\partial h_i(t_w)$ is the response  to an  
external magnetic field $h_i(t)$ in the $z$ direction and $X(t_w+t,t_w)$ is the 
fluctuation dissipation factor. In equilibrium the Fluctuation Dissipation Theorem (FDT) holds and T $X(t_w+t,t_w)= 1$, while out of 
equilibrium $X$ depends  on $t$ and $t_w$ in an non trivial way. It has been conjectured \cite{Cugliandolo93PRL} that 
 $X(t_w+t,t_w)= X[C(t_w+t,t_w)]$. This conjecture has proved valid in all systems studied to date. 

Instead of considering the response function it is easier to analyze 
the integrated response function 
\begin{equation}
\chi(t_w+t,t_w)=\int_{t_w}^{t_w+t} R(t_w+t,s)\,  ds.
\end{equation}
Assuming $X(t_w+t,t_w)= X[C(t_w+t,t_w)]$ one obtains
\begin{equation}
3 \, T\chi(t_w+t,t_w)= \int_{C(t_w+t,t_w)}^1 X(C) dC 
\end{equation}
and by plotting $3\, T\, \chi$ vs. $C$ one can extract $X$ from the slope of the curve.
In particular, if the FDT holds $X=1$ and $3 \, T\chi(t) = (1-C(t))$; any departure from 
this straight line  brings information about the non-equilibrium process.  In numerical simulations of spin glass 
\cite{Parisi98PRB} and structural glass models \cite{Kob97PRL,Parisi97PRL} it has 
been found that, in the non-equilibrium regime, this curve follows another straight 
line with smaller (in absolute value) slope when $t/t_w \gg 1$. In this case the FD factor $X$ can 
be interpreted as an effective inverse temperature \cite{Cugliandolo97PRE} $T_{eff}=T/X$. 
At time $t_w$ we took a copy of the system spin configuration, to which a random 
magnetic field $h_i(t)=h\, \epsilon_i$ was applied, in order to avoid favoring the 
QLRO \cite{Barrat98PRE,Stariolo99PRB}; $\epsilon_i$ was taken from a bimodal distribution 
($\epsilon_i=\pm 1$). Using the results from the FC and ZFC calculations, the strength 
$h$ of the field was taken small enough to guarantee linear response; the 
integrated response then equals $\chi(t_w+t,t_w)=m(t_w+t,t_w)/h$, where $m(t_w+t,t_w)$ is 
the staggered magnetization conjugated to the field $h_i(t)$, averaged over the random field variables.

\begin{figure}
\includegraphics*[width=8cm,angle=0]{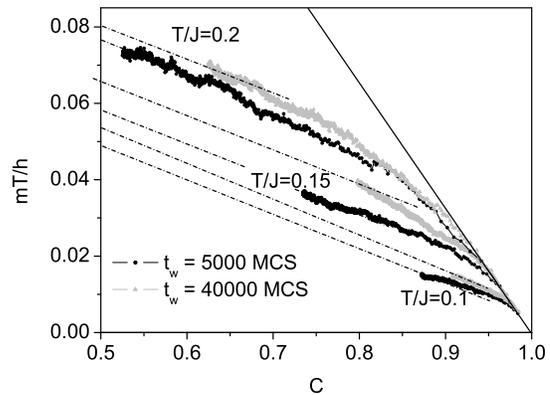}
\caption{\label{fig4} Parametric plot of $T\, m(t+t_{w}, t_w)/h$ versus $C(t+t_{w}, t_w)$ for three reduced 
temperatures $T/J$ ($0.1$, $0.15$ and $0.2$ from bottom to top) and two 
different waiting times $t_w$ ($5000$ and $40000$ MCS). The solid line 
indicates the validity of Fluctuation Dissipation Theorem (slope $1/3$).}
\end{figure}

In Fig. \ref{fig4} we display $T\, m(t_w+t,t_w)/h$ vs. $C(t_w+t,t_w)$ in a parametric plot. 
The curves correspond to the three different values of $T/J$ plotted in Fig. \ref{fig3}; in
each case we present the results obtained for two different
waiting times $t_w = 5000$ and $40000$ MCS. 
In all the cases studied we observe a typical two time scale separation
behavior, proper of real spin glasses. At $t=0$ the system starts
in the right bottom corner (fully correlated and demagnetized)
and during certain time (that depends in this case both on $t_w$
and $\mu$) it runs over the equilibrium straight line, indicating
the existence of a quasi equilibrium regime. Nevertheless, at
certain time the system clearly departures from this quasi--equilibrium
curve and moves along a different straight line, but with a different
(smaller) slope, indicating an effective temperature that is larger
than the temperature of the thermal bath.
This one--step temporal regime observed in this model is very common
in structural glasses but also in spin glass materials. Both
numerical results on the Heisenberg spin glass with weak anisotropy
\cite{Kawamura03PRL} and experimental measurements in 
CdIn$_{0.3}$Cr$_{1.7}$S$_4$ spin glasses, present this kind of
dynamical behavior.

\section{Conclusions}
In this paper we have studied the out of equilibrium
dynamics of the RAM model. The parameters have been chosen in such a way to model systems with
a freezing temperature well below the ordering temperature. 
We have restricted ourselves to consider the case of intermediate
values of $D/J$, where both effects exchange and randomness, 
compete with each other. This peculiar regime is specially interesting, since
on one hand there exists certain degree of controversy about the
expected ordering of the system, and on the other hand it allows
to model different interesting real magnetic materials. 
Our analysis, based the study of ZFC-FC curves, aging and on the FDR  in the low temperature phase, are consistent
with the existence of a spin--glass ground state of the model,
where the slowing mechanisms are then related to the topology of the energy landscape of the model.
  
This work was partially supported by grants from CONICET
(Argentina), Agencia C\'ordoba Ciencia (Argentina) and
SECYT/UNC (Argentina).

\end{document}